\documentclass[aps,prd,groupedaddress,amsmath,amssymb,amsfonts,nofootinbib,preprint]{revtex4-1}
\usepackage{amssymb,amsfonts}
\usepackage{mathrsfs}
\usepackage{tikz}
\usetikzlibrary{external}
\usepackage{amsthm}
\usepackage{xcolor}
\usepackage{graphicx}
\usepackage[utf8x]{inputenc}
\usepackage{float}
\usepackage{amsmath}
\usepackage{mathtools}
\usepackage{bbold}
\usepackage{accents}
\usepackage{cancel}
\DeclareMathAlphabet{\mathpzc}{OT1}{pzc}{m}{it}
\usetikzlibrary{patterns}

\usepackage{cleveref}

\begin{document}
\count \footins = 600

\title{Spin Self-Force}

\author{Kristian Mackewicz }
\email{kmackewicz@uchicago.edu}
\author{Robert M. Wald}
\email{rmwa@uchicago.edu}
\affiliation{Enrico Fermi Institute and Department of Physics \\
  The University of Chicago \\
  5640 South Ellis Avenue, Chicago, Illinois 60637, USA.}

\begin{abstract}

We consider the motion of charged and spinning bodies on the symmetry axis of a non-extremal Kerr-Newman black hole. If one treats the body as a test point particle of mass, $m$, charge $q$, and spin $S$, then by dropping the body into the black hole from sufficiently near the horizon, the first order area increase, $\delta A$, of the black hole can be made arbitrarily small, i.e., the process can be done in a ``reversible'' manner with regard to the change of parameters of the black hole. At second order, there may be effects on the energy delivered to the black hole---quadratic in $q$ and $S$---resulting from (i) the finite size of the body and (ii) self-force corrections to the energy. Sorce and Wald have calculated these effects for a charged, non-spinning body on the symmetry axis of an uncharged Kerr black hole and showed that, when these effects are included, $\delta^2 A$ also can be made arbitrarily small, i.e., this process remains reversible to second order. We consider the generalization of this process for a charged and spinning body on the symmetry axis of a Kerr-Newman black hole, where the self-force effects have not been calculated. A spinning body (with negligible extent along the spin axis) must have a mass quadrupole moment $\gtrsim S^2/m$, so at quadratic order in the spin, we must take into account quadrupole effects on the motion. After taking into account all such finite size effects, we find that the condition $\delta^2 A \geq 0$ yields a nontrivial lower bound on the self-force energy, $E_{SF}$, at the horizon. In particular, for an uncharged, spinning body on the axis of a Kerr black hole of mass $M$, the net contribution of spin self-force to the energy of the body at the horizon is $E_{SF} \geq S^2/8M^3$, corresponding to an overall repulsive spin self-force. A lower bound for the self-force energy, $E_{SF}$, for a body with both charge and spin at the horizon of a Kerr-Newman black hole is given. This lower bound will be the correct formula for $E_{SF}$ if the dropping process can be done reversibly to second order.
\end{abstract}

\maketitle

\tableofcontents

\section{Introduction}

It is commonly asserted that freely falling bodies in general relativity follow geodesics of the spacetime metric. However, this result is true only in the limit that both the size and mass of the body become arbitrarily small. If the body has finite size, $R$, then it can have a spin, $S$, as well as higher multipole moments, which will result in deviations from geodesic motion \cite{pap}, \cite{dix}. In addition, if the body has finite mass, $m$, it will perturb the background spacetime metric, also resulting in deviations from geodesic motion in the background spacetime. These deviations can be described as due to a ``self-force.''
Similarly, for a charged body, there will be deviations from Lorentz force motion caused by the electromagnetic multipole moments of the body arising from its finite size, and there will be self-force effects arising from its finite charge. For a charged body in curved spacetime, there will be electromagnetic self-force effects even if there is no external electromagnetic field.

If one neglects the effects of the body on the background spacetime metric and electromagnetic field---thereby treating the body as a ``test body'' and ignoring self-force effects---the effects of finite size on the motion can be analyzed in a systematic fashion. One first must define a suitable notion of a ``center of mass worldline'' to describe the motion of the body. This can be done in a satisfactory manner for sufficiently small bodies \cite{dix2}, \cite{schat1}, \cite{schat2}. At leading order in the size, $R$, of the body, there will be deviations from Lorentz force motion due to spin \cite{pap} and to electromagnetic dipole effects. At higher orders in the size of the body, higher gravitational and electromagnetic multipole moments will contribute. These all can be calculated systematically \cite{dix} (see also section 2.2 of \cite{har}).

It is much more difficult to analyze self-force effects.
Expressions for self-force can be rigorously derived by considering a suitable limit where the size, $R$, mass, $m$, and charge, $q$, of the body all scale to zero in proportional manner \cite{gw}. In this approach, one finds that the leading (``$0$th-order'') motion is given by Lorentz force motion. At the next order, the deviations from Lorentz force motion are given by the leading order self-force effects (i.e., force expressions quadratic in $m$ and $q$) as well as the test body spin and electromagnetic dipole terms \cite{gw}, \cite{ghw}. General expressions have been derived for the gravitational and electromagnetic self-force in a vacuum background spacetime, and progress has been made toward getting an explicit self-force expression in electrovac spacetimes \cite{zp}, where the coupled Einstein-Maxwell equations are needed. Second-order calculations (i.e., expressions cubic in $m$) of self-force have been given for uncharged bodies with no spin or quadrupole moment in a vacuum background spacetime \cite{pou1}, \cite{gra}, \cite{pou2}. However, the analysis of second order self-force is extremely complex, and it would not seem feasible at present to carry this type of analysis of self-force beyond second order.

It would be reasonable to expect that there should be self-force effects resulting from the spin, $S$, of the body (as well as from higher multipole moments). Specifically, the presence of spin will perturb the metric, and the effects of the perturbed metric back on the spin should perturb the motion. Such a ``spin self-force'' should be proportional to $S^2$ and thus should appear at third order in the perturbative scheme of the previous paragraph\footnote{There should also, in general, be self-force effects proportional to $mS$, which would appear at second order in the above perturbative scheme. However, our analysis will concern only the integrated self-force energy involved in lowering a spinning body to the horizon of a black hole along the symmetry axis. A term in the integrated self-force energy of the form $mS$ could take either sign depending on the orientation of the spin and cannot be present if the lowering process can be done reversibly when $S=0$, since by choosing the appropriate orientation, one could decrease the area of the black hole. In any case, we will consider only the self-force effects proportional to $S^2$.}. 
As an issue of principle, it would be of interest to calculate the effects of spin self-force. Although these effects presumably would be too small to have a significant effect on, e.g., the orbits of extreme mass ratio inspirals, they could be of importance for avoiding over-spinning of black holes \cite{js}. However, a direct calculation of spin self-force does not seem feasible.

In this paper, we will present indirect evidence for the existence of a spin self-force and will give a nontrivial lower bound for the ``spin self-force energy'' of a spinning body on the symmetry axis of a Kerr-Newman black hole at the horizon. We will do so by reversing the logic of a calculation done in the Appendix of a paper of Sorce and Wald \cite{sw}. Similar ideas were previously proposed by Hod \cite{hod} for the case of a charged particle with orbital angular momentum. Sorce and Wald considered lowering a small body of charge $q$ to the horizon of a non-extremal Kerr black hole along its symmetry axis and then dropping it in. At linear order in $q$, the body can be dropped arbitrarily close to the horizon and no energy is delivered to the black hole. At this order in $q$, there is no change in the mass and area of the black hole, i.e., $\delta M=0$ and $\delta A = 0$. At second order in $q$, the change in the black hole charge $\delta Q = q$ will contribute to a change, $\delta^2 A$, of the area of the black hole. However, at order $q^2$, there are two effects that give rise to a change $\delta^2 M$ in the mass of the black hole: (i) Assuming that the body is spherical in shape, the electromagnetic self-energy contribution to its mass is $\geq q^2/2R$, where $R$ is its proper radius. But one cannot lower such a body closer than a proper distance $R$ from the horizon. It follows that an energy of at least \cite{sw}
\begin{equation}
E_F = \frac{1}{2} \kappa q^2
\label{ep1}
\end{equation}
must be delivered to the black hole, where $\kappa$ is the surface gravity of the black hole. (ii) Charge self-force effects alter the energy delivered to the black hole. Specifically, work must be done by an external agent in overcoming the self-force when slowly lowering the body to near the horizon of the black hole. Since the charge self-force is repulsive, the external agent delivers an additional energy to the body (and ultimately to the black hole) given by
\begin{equation}
E_{SF}  = \int_{r_+}^\infty f(r) \sqrt{-g_{tt}} \sqrt{g_{rr}} dr \, ,
\label{sfe}
\end{equation}
where $r_+ = M + \sqrt{M^2 - a^2}$ is the radial coordinate of the horizon and $f(r)$ is the locally measured charge self-force on a static charged body (with the sign of $f$ chosen so that $f>0$ for a repulsive force). For the case of a Kerr black hole---where $f(r)$ is known---$E_{SF}$ is evaluated to be \cite{sw}
\begin{equation}
E_{SF}  = \frac{q^2}{4 r_+} \, .
\label{sfek}
\end{equation}
Sorce and Wald \cite{sw} showed that if the second order change of mass is given by $ E_F + E_{SF}$ with $E_F$ given by \cref{ep1} and $E_{SF}$ given by \cref{sfek}, we have $\delta^2 A = 0$. In other words, by dropping the charge as close as one can to the horizon, the area change of the black hole can be made to vanish to second order. Thus, the process of adding a charge to a Kerr black hole can be done ``reversibly'' \cite{chr} to second order.

Sorce and Wald \cite{sw} used the known expression for charge self-force to {\em prove} that the process of dropping a charged body with no spin into a non-extremal Kerr black hole from the horizon at the symmetry axis is reversible, i.e., is such that $\delta A = \delta^2 A = 0$. As we will show in sections \ref{csfe}, \ref{ssfe}, and \ref{tsfe} below, the process of dropping a charged and spinning body into a non-extremal Kerr-Newman black hole from the horizon at the symmetry axis is reversible at first order, $\delta A = 0$. However, self-force effects have not been calculated (except in the above case of a charged, non-spinning body on the axis of a Kerr black hole), so we cannot compute $E_{SF}$, as needed to compute $\delta^2 A$. Nevertheless, assuming that non-extremal Kerr-Newman black holes are stable, we know that if $\delta A = 0$, then we must have $\delta^2 A \geq 0$. We will show that this gives rise to a nontrivial lower bound on $E_{SF}$. If, in fact, the process is reversible to second order, then this lower bound will yield the exact expression for $E_{SF}$.

An important issue that we have to consider in our analysis is whether there is any analog of the energy contribution $E_F$, \cref{ep1}, for a spinning body that arises from its finite size. As we shall see in \cref{quad} below, a spinning body must have a ``size'' $R \geq 2 S/m$. However, in this case, $R$ represents the radius of a ring in a plane perpendicular to the symmetry axis; the size along the symmetry axis can be made arbitrarily small. Thus, the finite size of the spinning body does not provide an obstacle to lowering the body arbitrarily close to the black hole. However, such a ring will have a mass quadrupole moment of order $S^2/m$. This mass quadrupole moment could potentially affect the motion and the energy of the body at order $S^2$. Thus, it will be necessary for us to analyze quadrupole effects. We will see that, in fact, quadrupole effects do not prevent us from carrying out the dropping process nor do they contribute to the energy delivered to the black hole. Thus, only the spin self-force makes a contribution to the energy at order $S^2$, and the requirement that $\delta^2 A \geq 0$ then places a lower bound on the spin self-force energy.

In \cref{quad}, we analyze the motion of a charged and spinning test body on the symmetry axis of a stationary, axisymmetric, electrovac spacetime. In addition to charge, mass, and spin, the body is allowed to have a mass quadrupole moment, but it is assumed that all other electromagnetic and gravitational multipole moments vanish. After reviewing properties of the Kerr-Newman metric in \cref{kn},
we illustrate our method in \cref{csfe} by considering the dropping of a nonspinning charged body into a Kerr-Newman black hole from just outside the horizon on the symmetry axis. Our analysis runs in parallel with that of the Appendix of \cite{sw}
except that the self-force is not known in this case. Thus, instead of calculating $E_{SF}$ and showing that it yields $\delta^2 A = 0$, we assume that $\delta^2 A \geq 0$ and use this to obtain the lower bound \cref{esf1} for $E_{SF}$. We then apply the same method in \cref{ssfe} to obtain the lower bound \cref{esf2} for the spin self-force energy of a spinning body on the horizon of an uncharged Kerr black hole. Finally, in \cref{tsfe} we obtain the lower bound \cref{totsf} for the total self-force energy in the general case of a charged and spinning body at the horizon of a Kerr-Newman black hole.

\section{Motion of a Charged, Spinning Test Body with a Quadrupole Moment on the Symmetry Axis of a Stationary, Axisymmetric Electrovac Spacetime}
\label{quad} 

Consider a body with stress-energy $T_{ab}$ and charge-current $j^a$ in an arbitrary curved spacetime with metric $g_{ab}$ in which an electromagnetic field $A_a$ is present. We treat the body as a test body, i.e., we neglect its effect on $g_{ab}$ and $A_a$ and thus do not include self-force effects. In the following sections, the failure to achieve $\delta^2 A \geq 0$ in black hole processes in the test body approximation will be used to deduce information about self-force. 

By conservation of stress-energy and charge-current, we have 
\begin{align}
\label{tcons}
    \nabla_b T^{ab}&=F^{ab} j_b \\
    \nabla_aj^a&=0 \, , 
\label{jcons}
\end{align}
where $F_{ab}=\nabla_a A_b-\nabla_b A_a$. Equations of motion for the body were derived from these relations by Dixon \cite{dix}. For the case of a body that has no gravitational multipole moments other than mass, spin, and mass quadrupole moment and has no electromagnetic multipole moments other than charge, the equations of motion are \cite{dix}
\begin{equation}
    \frac{Dp_a}{D\tau}=qv^b F_{ab} -\frac{1}{2}R_{abcd}v^bS^{cd} +\frac{1}{6}(u_bv^b)J^{cdef}\nabla_a R_{cdef}
\label{eomp}
\end{equation}
\begin{equation}
    \frac{DS^{ab}}{D\tau}=2p^{[a}v^{b]}+\frac{4}{3}(u_cv^c)J^{def[a}{R_{def}}^{b]} \, .
\label{eoms}
\end{equation}
Here $p^a$ is the $4$-momentum of the body, $S^{ab}$ is its spin tensor, $J^{abcd}$ is its mass quadrupole tensor (which also includes effects of momentum and stress), and $q$ is its charge. The spin tensor is antisymmetric, $S^{ab} = - S^{ba}$, and the quadrupole tensor has the symmetries of the Riemann tensor. The mass of the body is given by $m^2 = - p^a p_a$ and $u^a$ is defined by $u^a = p^a/m$. The quantity $v^a$ is the $4$-velocity of the center of mass worldline and $D/D\tau = v^a \nabla_a$. Note that $v^a$ need not be colinear with $p^a$. The spin tensor and 4-momentum satisfy the relation $p_a S^{ab} = 0$ as a consequence of the definition of the center of mass. We refer the reader to Dixon's paper \cite{dix} (see also section 2.2 of \cite{har}) for the definitions of $p^a$, $S^{ab}, J^{abcd}$, and the center of mass worldline, as well as for the derivations of \cref{eomp} and \cref{eoms}.
Note that there is no equation of motion for $J^{abcd}$; its evolution will depend on detailed properties of the body that go beyond the general conservation laws eqs.~(\ref{tcons}) and (\ref{jcons}).

Now suppose that the spacetime admits a Killing field $\chi^a$ that is also a symmetry of $A_a$ so that 
\begin{equation}
    \mathcal{L}_{\chi}g_{ab}=0 \, ,\quad \mathcal{L}_{\chi} A_a=0 \, .
\end{equation}
Then eqs.~(\ref{tcons}) and (\ref{jcons}) imply that 
\begin{equation}
    \nabla_b (T^{ab}\chi_a+A^a j^b \chi_a)=0 \, , 
\end{equation}
which, in turn, implies that the quantity
\begin{equation}
\alpha \equiv \int_{\Sigma} \sqrt{-g}(T^{ab}+A^a j^b)\chi_a d\Sigma_b \, ,
\label{kcons}
\end{equation}
is conserved, i.e., independent of the surface $\Sigma$ over which the integral in \cref{kcons} is taken. This conserved quantity can be expressed in terms of the $4$-momentum, charge, and spin tensor by \cite{dix}
\begin{equation}
    \alpha= -(p_a+qA_a)\chi^a - \frac{1}{2}S^{ab}\nabla_a \chi_b \, .
\end{equation}
Note that the quadrupole moment tensor does not enter the expression for $\alpha$ (nor would other gravitational or electromagnetic multipole moments enter this expression if they were non-zero). One may directly verify the constancy of $\alpha$ from the equations of motion (\ref{eomp})-(\ref{eoms}) by the following calculation

\begin{align}
    \frac{D \alpha}{D \tau}&= -(\frac{Dp_a}{D\tau}+q\frac{D A_a}{D\tau})\chi^a -(p_a+qA_a)v^b \nabla_b \chi^a -\frac{1}{2}\frac{D S^{ab}}{D\tau}\nabla_a\chi_b -\frac{1}{2}S^{ab}v^c \nabla_c \nabla_a \chi_b \\ \nonumber
    &=\frac{1}{6}\chi^aJ^{bcde}\nabla_aR_{bcde} -q\chi^a v^b F_{ab}  - q \chi^av^b \nabla_b A_a - q A_a v^b \nabla_b \chi^a+\frac{2}{3}J^{cde[a}{R_{cde}}^{b]}\nabla_a \chi_b \\ \nonumber
    & = \frac{1}{6}J^{abcd}\mathcal{L}_{\chi}R_{abcd}-v^a\mathcal{L}_{\chi}A_a \\ \nonumber
    &=0 \, ,
\end{align}
where $\nabla_c\nabla_a \chi_b=-R_{abc}^{\quad d}\chi_d$ was used in the second line.

Now consider a stationary and axisymmetric spacetime with timelike (at infinity) Killing field $\xi^a$ and axial Killing field $\psi^a$. Then the corresponding conserved quantities are naturally referred to as the {\em energy}, $\epsilon$, and {\em angular momentum}, $j$, of the test body
\begin{equation}
    \epsilon \equiv -(p_a+qA_a)\xi^a - \frac{1}{2}S^{ab}\nabla_a \xi_b
\label{endef}
\end{equation}
\begin{equation}
    j \equiv (p_a+qA_a)\psi^a + \frac{1}{2}S^{ab}\nabla_a \psi_b \, .
\label{amdef}
\end{equation}

Despite the presence of these constants of motion, the analysis of solutions to \cref{eomp} and \cref{eoms} in a stationary, axisymmetric spacetime can be quite complex. However, the analysis simplifies significantly if we restrict attention to axisymmetric test bodies, $\mathcal{L}_{\psi}T_{ab} = \mathcal{L}_{\psi}j_a = 0$. The center of mass of such a body must lie on the symmetry axis, $\psi^a = 0$. The ``symmetry axis'' is a two dimensional timelike surface, so the motion in this case is in only one spatial dimension. It is convenient to introduce an orthonormal basis $\hat{t}^a, \hat{z}^a, \hat{x}^a, \hat{y}^a$ at each event on the symmetry axis as follows: We choose $\hat{t}^a = \xi^a/(-\xi^b \xi_b)^{-1/2}$. We choose $\hat{z}^a$ to lie tangent to the symmetry axis (and orthogonal to $\hat{t}^a$), and we choose $\hat{x}^a$ and $\hat{y}^a$ to be orthogonal to the symmetry axis, with the $\hat{x}^a, \hat{y}^a, \hat{z}^a$ chosen to have positive orientation and be such that $\nabla_a \psi_b = \hat{x}^a \hat{y}^b - \hat{y}^a \hat{x}^b$. Since, by symmetry, $p^a$ must be tangent to the symmetry axis, and since $p_a S^{ab} = 0$, it follows that $S^{ab}$ takes the form
\begin{equation}
S^{ab} = S (\hat{x}^a \hat{y}^b - \hat{y}^a \hat{x}^b) \, ,
\label{s1}
\end{equation}
where we refer to $S$ as the spin of the body. The conserved quantity $j$ of \cref{amdef} is then simply
\begin{equation}
    j = S \, ,
\end{equation}
so the spin is conserved. 

We now make the further assumption that the background spacetime metric is ``electrovac,'' i.e., a solution to the Einstein-Maxwell equations. As shown in the Appendix, this implies that on the symmetry axis we have $R_{ab} l^a l^b = R_{ab} n^a n^b = 0$, where $l^a = (\hat{t}^a + \hat{z}^a)/\sqrt{2}$, $n^a = (\hat{t}^a - \hat{z}^a)/\sqrt{2}$. From the general form of the Riemann tensor on the symmetry axis, it is shown in the Appendix that the ``torque term," $J^{def[a}{R_{def}}^{b]}$, in \cref{eoms} vanishes on the symmetry axis. Since $D S^{ab}/D \tau$ also vanishes, it follows from \cref{eoms} that $p^a$ and $v^a$ are colinear, so $p^a = m v^a$. 

It remains to solve \cref{eomp}. The analysis of this equation can be greatly simplified by using the constant of motion \cref{endef}. Let $t,z$ be coordinates on the symmetry axis such that $(\partial/\partial t)^a= \xi^a$ and $(\partial/\partial z)^a = \hat{z}^a$, i.e., $t$ is a Killing coordinate and $z$ is a proper distance coordinate on the symmetry axis. 
Since $p^a = m v^a$ we have
\begin{equation}
p_a \xi^a =  m g_{tt} \dot{t} \, ,
\end{equation}
where the overdot denotes $d/d\tau$, where $\tau$ is the proper time along the center of mass worldline. Using \cref{s1}, we have
\begin{equation}
S^{ab}\nabla_a \xi_b = \beta(z) S \, ,
\end{equation}
where 
\begin{equation}
\beta(z) \equiv (\nabla^a \psi^b) \nabla_a \xi_b \, .
\end{equation}
Thus, we obtain
\begin{equation}
    \epsilon =- m  g_{tt} \dot{t} +q \Phi - \frac{1}{2} \beta S \, ,
\label{en2}
\end{equation}
where we have written $\Phi = - A_a \xi^a$. \Cref{en2} can be used to solve for $\dot{t}$
\begin{equation}
   \dot{t} = \frac{1}{m (-g_{tt})} \left[\epsilon - q \Phi + \frac{1}{2} \beta S \right] \, .
\label{tdot}
\end{equation}
This can then be substituted into the equation
\begin{equation}
-1 = v^a v_a = g_{tt} \dot{t}^2 + \dot{z}^2
\end{equation}
to obtain
\begin{equation}
\dot{z}^2 = \frac{1}{m^2 (-g_{tt})}\left[\epsilon - q \Phi + \frac{1}{2} \beta S \right]^2 - 1 \, .
\label{zevol}
\end{equation}
In this equation, $g_{tt}$ and $\beta$ are known functions of $z$ determined by the background spacetime metric, and $\epsilon$ and $S$ are constants of motion, determined by the initial conditions of the body. 

By conservation of charge, $q$ also is a constant of motion. However, $m$ is {\em not} in general a constant of motion. Indeed, its evolution is determined by contracting $p^a$ into \cref{eomp}. We obtain
\begin{equation}
\frac{dm^2}{d\tau}=-2 p^a  \frac{Dp_a}{D\tau} =  \frac{1}{3}mJ^{abcd}\frac{D}{D\tau}R_{abcd} \, .
\label{mevol}
\end{equation} 
Thus, if the quadrupole moment is nonvanishing, we must solve \cref{mevol} along with \cref{zevol}. In order to solve \cref{mevol}, we must know the evolution of $J^{abcd}$, which is not determined from the conservation relation (\ref{tcons}) alone. In \cref{ssfe} below, we will solve \cref{mevol} in Kerr for the case of a spinning ring of minimal radius for its spin.

Note that the quadrupole moment tensor does not enter \cref{zevol}. Thus, the quadrupole moment of the body affects its motion along the symmetry axis only by affecting its rest mass via \cref{mevol}. This is not how quadrupole effects are normally described in Newtonian gravity, so it is worthwhile explicitly seeing how Newtonian behavior arises for an uncharged body undergoing nonrelativistic motion along the $z$-axis in a nearly Newtonian spacetime\footnote{The Newtonian behavior of such a body also could be derived directly from \cref{eomp}. The aim of the calculation below is to show how \cref{zevol}---wherein the quadrupole effects enter solely via changes to the rest mass---is compatible with Newtonian behavior.}.

In a nearly Newtonian spacetime, we have
\begin{equation}
g_{tt} \approx - (1 + 2 \phi) \, ,\quad \quad \beta \approx 0 \, ,
\end{equation}
where $\phi$ is the Newtonian potential. By assumption, we have $|\phi| \ll 1$ and time derivatives of $\phi$ are negligibly small. For a nonrelativistic body, we have
\begin{equation}
J^{abcd} \approx \frac{3}{4}\left(\Hat{t}^a\Hat{t}^c H^{bd} -\Hat{t}^b\Hat{t}^c H^{ad}-\Hat{t}^a\Hat{t}^d H^{bc}+\Hat{t}^b\Hat{t}^d H^{ac} \right) \, ,
\end{equation}
where $H^{ab}$ has purely spatial components, given by
\begin{equation}
    H^{ij}=\int x^i x^j \rho(x) d^3x \, . 
\end{equation}
To linear order in $\phi$, the Riemann tensor is given by
\begin{equation}
    R_{abcd} \approx \Hat{t}_b \Hat{t}_d \partial_a \partial_c \phi -\Hat{t}_a \Hat{t}_d \partial_b \partial_c \phi -\Hat{t}_b \Hat{t}_c \partial_a \partial_d \phi + \Hat{t}_a \Hat{t}_c \partial_b \partial_d \phi \, .
\end{equation}
To linear order in $\phi$, \cref{zevol} yields
\begin{equation}
\dot{z}^2 = \frac{\epsilon^2}{m^2} (1 - 2 \phi) - 1 \, .
\label{zevoln}
\end{equation}
\Cref{mevol} yields
\begin{align}
    \dot{m} &= \frac{1}{6} J^{\mu \nu \lambda \rho} \partial_z (R_{\mu \nu \lambda \rho}) \dot{z} \nonumber \\
    &= \frac{1}{2} H^{\mu \nu} (\partial_z \partial_\mu \partial_\nu  \phi) \dot{z} \, .
\label{mdot}
\end{align}
Differentiating \cref{zevoln} with respect to $\tau$ and, again, neglecting higher order terms in $\phi$, we obtain
\begin{equation}
2 \dot{z} \ddot{z} = -\frac{\epsilon^2}{m^2} 2 (\partial_z \phi) \dot{z} - 2 \frac{\epsilon^2}{m^3} \dot{m} \, .
\label{zevoln2}
\end{equation}
We now may substitute \cref{mdot} for $\dot{m}$.
For non-relativistic motion, we can ignore the difference between proper time $\tau$ and coordinate time $t$, i.e., we may set $\dot{t} \approx 1$. We also may approximate $\epsilon \approx m$ on the right side of \cref{zevoln2} 
since, by \cref{en2}, $(\epsilon - m)$ is of order $\phi$. Thus, we obtain
\begin{equation}
m \frac{d^2 z}{dt^2} = - m \partial_z \phi  -  \frac{1}{2} H^{\mu \nu} (\partial_z \partial_\mu \partial_\nu  \phi) \, .
\label{zevoln3}
\end{equation}
This is exactly what Newton would have expected. Of course, Newton would have attributed the first term on the right side as being due to a ``gravitational force'' rather than geodesic motion in a curved spacetime, and he would not have associated the quadrupole force with the change in the rest mass of the body.

\section{The Kerr-Newman Black Hole and its First and Second Order Area Variations}
\label{kn}

The Kerr-Newman metric of mass $M$, charge $Q$, and angular momentum $J=aM$ is given in Boyer-Lindquist coordinates by
 \begin{align}
     ds^2&= -\frac{\Delta -a^2 \sin^2\theta}{\Sigma}dt^2 - \frac{2a \sin^2\theta(r^2+a^2-\Delta)}{\Sigma}dtd\phi \\ \nonumber
    &+ \frac{\Sigma}{\Delta}dr^2+\Sigma d\theta^2 + \frac{(r^2+a^2)^2-\Delta a^2 \sin^2\theta}{\Sigma}\sin^2\theta d\phi^2 \, ,
 \end{align}
 where
 \begin{align}
     \Delta&=r^2+a^2+Q^2-2 M r \\
     \Sigma&=r^2+a^2 \cos^2\theta \, .
 \end{align}
 The vector potential is given by
 \begin{equation}
     A_adx^a=-\frac{Q r}{\Sigma}(dt-a \sin^2 \theta d\phi) \, .
 \end{equation}
 These solutions describe black holes when $Q^2 + a^2 \leq M^2$.
The horizon of the black hole is located at 
 \begin{equation}
     r=r_+=M+\sqrt{M^2-a^2-Q^2} \,.
 \end{equation}
 The area of the horizon is given by
 \begin{equation}
    A= 4 \pi (r_+ ^2 +a^2) \,.
\end{equation}
The angular velocity of the horizon is given by
 \begin{equation}
    \Omega_H = \frac{a}{r_+^2 + a^2} \,.
\end{equation}
The horizon Killing field is $\chi^a = \xi^a + \Omega_H \psi^a$. We define $\Phi = -A_a \chi^a$.
The electrostatic potential at the horizon is given by
 \begin{equation}
    \Phi_H \equiv \Phi(r_+) =  \frac{Q r_+}{r_+^2 + a^2} \, .
\end{equation}
The surface gravity of the horizon is
 \begin{equation}
   \kappa = \frac{r_+ - M}{r_+^2 + a^2} \, .
\label{kappa}
\end{equation}
We restrict consideration in this paper to the non-extremal case, $\kappa > 0$, i.e., $Q^2 + a^2 < M^2$

In this paper, we will be concerned with the first and second order area changes of the black hole when a charged, spinning body is dropped into the black hole. Thus, we consider one parameter families of metrics $g_{ab}(\lambda)$, stress-energies $T_{ab}(\lambda)$, and charge-currents $j^a(\lambda)$, corresponding to spacetimes where, initially, one has a Kerr-Newman black hole of parameters $(M, Q, J)$ with charged matter outside the black hole; one then drops the charged matter into the black hole and lets the black hole settle down to a final state Kerr-Newman black hole with parameters $(M(\lambda), Q(\lambda), J(\lambda))$. One can choose the stress-energy and charge-current of the matter so that the matter carries no charge and angular momentum beyond first order, so that we have
 \begin{align}
    Q(\lambda)&=Q+\lambda \delta Q \\
    J(\lambda)&=J+\lambda \delta J \, .
\end{align}
where for any quantity $X$ we write
\begin{equation}
    \delta X = (\frac{\partial X}{\partial \lambda})|_{\lambda=0} \, , \quad \delta^2 X = (\frac{\partial^2 X}{\partial \lambda^2})|_{\lambda=0} \, , \dots
\label{d2}
\end{equation}
As stated above, we take $\delta^2 Q = \delta^2 J = 0$.
However, the second order change in mass of the black hole will be affected by the finite size and self-force energy of the body, so we cannot assume that $\delta^2 M$ vanishes. 

The first order change in the area of the final state Kerr-Newman black hole is given by the first law of black hole mechanics
\begin{equation}
    \frac{\kappa}{8\pi}\delta A=\delta M-\Omega_H \delta J-\Phi_H \delta Q \,.
\label{da}
\end{equation}
Below, we will primarily be concerned with the case where $\delta A = 0$. In this case, using $\delta^2 Q = \delta^2 J = 0$, the second order variation of area will be given by
\begin{equation}
    \frac{\kappa}{8\pi}\delta^2 A = \delta^2 M -\delta (\Omega_H)\delta J - \delta(\Phi_H)\delta Q \, ,
\label{d2a}
\end{equation}
i.e., it will have a contribution from $\delta^2 M$ and from terms quadratic in the first order variations. Since $\delta A = 0$, we can express $\delta M$ in terms of $\delta Q$ and $\delta J$ by \cref{da}, so the quadratic terms in \cref{d2a} can be written purely in terms of $\delta Q$ and $\delta J$.

\section{Self-Force Energy for a Charged Body in Kerr-Newman}
\label{csfe}

In this section we illustrate our method by considering a body of charge $q$, and mass $m$---but no spin or higher multipole moments---on the symmetry axis of a Kerr-Newman black hole. Sorce and Wald \cite{sw} treated the case of an uncharged Kerr black hole---where the charge self-force is known---and showed that the process can be done reversibly to second order, $\delta A = \delta^2 A = 0$, if the charge is dropped as near as possible to the horizon. Here, we consider the case of a Kerr-Newman black hole, where the charge self-force is not known. We will use the fact that we must have $\delta^2 A \geq 0$ when $\delta A = 0$ to deduce a lower bound for the charge self-force energy at the horizon. This lower bound will yield the exact expression for the self-force energy if the process is, in fact, reversible to second order. 

Consider a body on the symmetry axis of a Kerr-Newman black hole with mass $m$ and charge $q$ but with $S=0$ and $J^{abcd} = 0$. We initially treat the body as a test body of arbitrarily small size. By \cref{mevol}, $m$ is constant. The energy is given by
\begin{equation}
    \epsilon =- m  g_{tt} \dot{t} +q \Phi \, .
\label{en3}
\end{equation}
The motion of the body is determined by \cref{zevol}. Replacing the proper distance coordinate $z$ by the Boyer-Lindquist radial coordinate $r$, we obtain
\begin{equation}
    m^2 \Dot{r}^2+ V_{\rm eff}(r)=0 \, ,
\end{equation}
where
\begin{equation}
    V_{\rm eff}(r)=-(\epsilon - q \frac{Q r}{r^2+a^2})^2+m^2(\frac{\Delta}{r^2+a^2}) \, .
\end{equation}
We now drop the body from rest at $r=r_0$. Since $\dot{r} = 0$ at $r=r_0$, we have 
\begin{equation}
V_{\rm eff}(r_0)=0 \, ,
\end{equation}
which can be used to solve for $\epsilon$ in terms of $m$, $q$, and $r_0$
\begin{equation}
  \epsilon = q \frac{Q r_0}{r_0^2+a^2}+m \left(\frac{\Delta(r_0)}{r_0^2+a^2} \right)^{1/2} \, .
\label{en4}
\end{equation}
In order that the body initially fall towards the black hole, we must have 
\begin{equation}
\frac{d V_{\rm eff}}{dr} (r_0) > 0 \,.
\end{equation}
This relation requires 
\begin{equation}
    m> q Q \frac{r_0^2-a^2}{M r_0^2-Q^2 r_0-a^2M}\sqrt{\frac{\Delta(r_0)}{r_0^2 + a^2}}  \,. 
\label{mbnd}
\end{equation}
For $r_0$ sufficiently close to the horizon, it can be checked that if \cref{mbnd} holds, then $V_{\rm eff}(r) < 0$ for $r_+ \leq r < r_0$, so there will be no turning points, and the body will go into the black hole.

From \cref{en4} and \cref{mbnd} we see that for the body to go into the black hole, we must have
\begin{equation}
  \epsilon > q \frac{Q r_0}{r_0^2+a^2}
\label{en5}
\end{equation}
with equality arbitrarily close to being achieved in the limit as the dropping point $r_0$ approaches $r_+$. If the body goes into the black hole, we have $\delta M = \epsilon$ and $\delta Q = q$. From \cref{da}, we have
\begin{equation}
    \frac{\kappa}{8\pi}\delta A=\delta M-\Phi_H \delta Q = \epsilon - q \frac{Q r_+}{r_+^2+a^2} \to 0
\label{da2}
\end{equation}
in the limit $r_0 \to r_+$.

At second order, we take $\delta^2 Q = 0$, i.e., we do not add additional charge to the black hole. As already discussed in the introduction, there are two effects that contribute to $\delta^2 M$. The first arises from the fact that the electromagnetic self-energy of the body provides a lower bound to the mass, $m \geq q^2/2R$, where $R$ is the proper radius of the body. However, a finite $R$ prevents one from lowering the body to closer than a proper distance $R$ from the horizon of the black hole. This finite size effect implies that an energy of at least\footnote{Here we have assumed that the charged body is spherical in shape. We could lower the body closer to the horizon by taking it to be smaller in one spatial dimension than the others, at only a modest cost in the self-energy, thereby decreasing $E_F$. However, the body would then acquire an electric quadrupole moment, which would contribute to the self-force. Since we are interested in the self-force associated with a ``pointlike,'' spherical body, we take the charge to be spherically distributed. In that case, the self-energy is minimized by distributing the charge as a spherical shell of radius $R$.}
\begin{equation}
E_F = \frac{1}{2} \kappa q^2
\label{fsq}
\end{equation}
must be delivered to the black hole \cite{sw}. 

The second effect is that of charge self-force, which contributes an energy $E_{SF}$. This energy is given by \cref{sfek} for the case of an Kerr black hole but is not known for the case of a charged Kerr-Newman black hole. The total second order change in mass\footnote{The factor of $1/2$ comes from the Taylor coefficient, which we chose not to incorporate into our definition of $\delta^2$ in \cref{d2}.}, $\frac{1}{2} \delta^2 M$, should therefore be given by
\begin{equation}
\frac{1}{2} \delta^2 M = E_F + E_{SF} \,.
\end{equation}
Substituting this expression in the formula for the second variation of area, we obtain
\begin{align}
   \frac{\kappa}{8 \pi} \delta^2 A &= \delta^2 M -\delta(\Phi_H)\delta Q \nonumber \\
   &= \delta^2 M- q(\delta Q \frac{\partial \Phi_H}{\partial Q}+ \delta M \frac{\partial \Phi_H}{\partial M}) \nonumber \\
   &=2(E_F + E_{SF}) -q^2 \frac{r_+}{r_+^2+a^2}-q^2\frac{a^2 Q^2}{M(r_+^2+a^2)^2} \nonumber \\
   &=2E_S -q^2 \frac{M}{r_+^2+a^2}-q^2\frac{a^2 Q^2}{M(r_+^2+a^2)^2} \, .
\end{align}

Thus, we see that in order that $\delta^2 A \geq 0$, we must have
\begin{equation}
    E_{SF} \geq \frac{q^2}{2M (r_+^2+a^2)^2}(M^2(r_+^2+a^2)+a^2Q^2) \, ,
\label{esf1}
\end{equation}
which gives a nontrivial lower bound on the self-force energy \cref{sfe}. When $Q = 0$, this lower bound yields the actual self-force energy \cref{sfek}.

\section{Spin Self-Force Energy in Kerr}
\label{ssfe}

In this section, we consider the case of an uncharged spinning body of mass $m$ and spin $S$ on the symmetry axis of an uncharged Kerr black hole that is dropped into the black hole from very near the horizon. The body is assumed to be axisymmetric, as analyzed in \cref{quad}.
On account of its spin, the body must have a finite extent $R \sim S/m$ perpendicular to the symmetry axis, so we will need to take quadrupole effects into account at order $S^2$.

As in the previous section, we initially treat the body as a test body. Converting the proper distance coordinate $z$ of \cref{zevol} to the Boyer-Lindquist coordinate $r$, we find that the motion is given by
\begin{equation}
    m^2 \Dot{r}^2+ V_{\rm eff}(r)=0 \, ,
\end{equation}
where
\begin{equation}
    V_{\rm eff}(r)=-(\epsilon - 2 M a S \frac{r}{r^2+a^2})^2 +m^2\frac{\Delta}{r^2 + a^2}
\label{veffs}
\end{equation}
At first order, we neglect the quadrupole moment, in which case $m$ is constant. If we drop the body from rest at $r = r_0$, we have $V_{\rm eff}(r_0) = 0$, which implies
\begin{equation}
    \epsilon= 2 M a S \frac{ r_0}{(r_0^2 + a^2)^2} + m \left(\frac{\Delta(r_0)}{r_0^2+a^2} \right)^{1/2} \, .
\label{en6}
\end{equation}
In order that the body initially fall towards the black hole, we must have 
\begin{equation}
\frac{d V_{\rm eff}}{dr} (r_0) > 0 \, .
\label{dv0}
\end{equation}
This relation requires 
\begin{equation}
 m > 2 a S \left(\frac{3r_0^2-a^2}{r_0^4-a^4}\right) \left(\frac{\Delta(r_0)}{r_0^2+a^2} \right)^{1/2} \, .
\label{mbnds}
\end{equation}
It can be checked that, for $r_0$ sufficiently close to $r_+$, if \cref{mbnds} holds, then $V_{\rm eff}(r) < 0$ for $r_+ \leq r < r_0$, so there will be no turning points, and the body will go into the black hole.

From \cref{en6} and \cref{mbnds} we see that for the body to go into the black hole, we have
\begin{equation}
    \epsilon > 2 M a S \frac{ r_0}{(r_0^2 + a^2)^2}  \, ,
\label{en7}
\end{equation}
with equality arbitrarily close to being achieved in the limit as the dropping point $r_0$ approaches $r_+$. If the body goes into the black hole, we have $\delta M = \epsilon$ and $\delta J = S$. From \cref{da}, we have
\begin{equation}
    \frac{\kappa}{8\pi}\delta A=\delta M-\Omega_H \delta J = \epsilon - \frac{a}{r_+^2 + a^2} S \, .
\label{da2s}
\end{equation}
Thus we see that $\delta A \to 0$ in the limit $r_0 \to r_+$, where we have used the fact that $r_+^2 + a^2 = 2Mr_+$ in Kerr. 

At second order, we take $\delta^2 J = 0$, but we have to take into account effects of order $S^2$ associated with (i) the finite size of the body and (ii) spin self-force corrections to the energy. We will be able to deduce a bound on the spin self-force energy if we can calculate the effects of finite size, so we turn our attention now to finite size effects.

We take as a model of a spinning body in flat spacetime in cylindrical coordinates 
$(t,\rho, \phi, z)$
\begin{equation}
    T^{ab}=\frac{m}{2 \pi R} k^{(a} t^{b)}\delta(z)\delta(\rho-R) \, ,
\end{equation}
where 
\begin{equation}
k^\mu=(1,0,\frac{1}{\rho},0) \, ,
\end{equation}
i.e., $k^a$ is a null vector in the $\phi$-direction.
This corresponds to having a ring of matter of cylindrical radius $R$ and negligible $z$-extent that is rotating as rapidly as possible consistent with stress-energy conservation and the dominant energy condition. Here $m$ is the mass of the body
\begin{equation}
m = \int T_{ab} t^a t^b d^3x \, .
\end{equation}
The angular momentum of the body is given by
\begin{equation}
S = - \int T_{ab} t^a \phi^b d^3x = \frac{1}{2} m R \, .
\label{spr}
\end{equation}
Using the definition of quadrupole moment tensor given in \cite{dix} specialized to flat spacetime, we find that the nonvanishing components of the mass quadrupole tensor are
\begin{equation}
    J^{xtxt}=J^{ytyt}=\frac{3}{8}m R^2 \, .
\label{quadr}
\end{equation}

If we put such a body on the symmetry axis of a Kerr black hole, the curvature corrections to the structure of the body will be negligible at the order to which we work. However, if we drop the body, then as we found in \cref{quad}, its mass will no longer be constant due to quadrupole effects. Since $S= m R/2$ is constant, the radius of the body also must change correspondingly. 
The nonvanishing components of the quadrupole moment tensor are given by \cref{quadr} in the body's rest frame. Thus, the quadrupole moment tensor is given by
\begin{equation}
    J^{abcd} =\frac{3}{2}m R^2 \left( \hat{x}^{[a} v^{b]} \hat{x}^{[c} v^{d]} + \hat{y}^{[a} v^{b]} \hat{y}^{[c} v^{d]} \right) \, .
\label{quadr2}
\end{equation}
Thus, using \cref{spr}, we have
\begin{equation}
    m J^{abcd} =6 S^2 \left( \hat{x}^{[a} v^{b]} \hat{x}^{[c} v^{d]} + \hat{y}^{[a} v^{b]} \hat{y}^{[c} v^{d]} \right) \, .
\label{quadr3}
\end{equation}
Since $v^a \nabla_a v^b = O(S)$, to order $S^2$ we have
\begin{equation}
\frac{D}{D \tau} (m J^{abcd}) =  v^e \nabla_e (m J^{abcd}) = 0 \, .
\end{equation}
Therefore, by \cref{mevol}, we have 
\begin{align}
    \frac{d m^2}{d \tau}&= \frac{1}{3}m J^{abcd} \frac{D}{D\tau}R_{abcd} \nonumber \\
    &= \frac{1}{3} \frac{d}{d\tau} \left(m J^{abcd} R_{abcd} \right) \nonumber \\
    &= 2 M S^2 \frac{d}{d\tau} \left( \frac{r(r^2-3a^2)}{(r^2+a^2)^3} \right) \, .
\end{align}
Thus, we find
\begin{equation}
    m(r)^2 = m_0^2 + 2 M S^2 \left(\frac{r(r^2-3a^2)}{(r^2+a^2)^3}-\frac{r_o(r_o^2-3a^2)}{(r_o^2+a^2)^3} \right) \, . 
\label{mevol2}
\end{equation}
where $m_0$ is the initial mass of the body when it is dropped from $r=r_0$. We may then plug \cref{mevol2} into \cref{veffs} to get the effective potential for the motion of the body including the quadrupole moment.

We now have the all of the results needed to analyze the finite size effects. Unlike the charged case analyzed in the previous section, the spinning ring need only have size $R$ in the directions perpendicular to the symmetry axis. Thus, there is no obstacle arising from the finite size of the body to lowering it arbitrarily close to the horizon. In addition, as discussed in \cref{quad}, there also is no quadrupole moment contribution to the energy. Nevertheless, the quadrupole moment affects the motion in the manner indicated in the previous paragraph. Specifically, it will alter $V_{\rm eff}$ and thereby affect the condition given by \cref{dv0}. The modification to $V_{\rm eff}$ also could potentially result in the existence of new turning points between $r_0$ and $r_+$, thereby possibly giving rise to additional restrictions for the body to enter the black hole. Finally, turning points could also result from $m^2$ going through $0$ in \cref{mevol2}. However, it is easily checked one can choose $m_0$ sufficiently large that (i) \cref{dv0} is satisfied, (ii) $V_{\rm eff}(r) < 0$ for $r_0 > r \geq r_+$, (iii) $m^2(r) > 0$ for $r_0 \geq r \geq r_+$, and such that (iv) $\epsilon$ approaches the lower bound \cref{en7} arbitrarily closely as $r_0 \to r_+$.
{\em Thus, we conclude that for a spinning body there are no finite size corrections to the energy at order $S^2$, i.e., we have}
\begin{equation}
E_F = 0 \, .
\end{equation}

Thus, the entire second order change in mass must be due to the self-force energy
\begin{equation}
\frac{1}{2} \delta^2 M = E_{SF} \, .
\end{equation}
The second order change in the area of the black hole is therefore given by
\begin{align}
    \frac{\kappa}{8 \pi}\delta^2 A &=\delta^2 M -\delta(\Omega_H)\delta J \nonumber \\
    &=\delta^2 M- S(\delta J \frac{\partial \Omega_H}{\partial J}+ \delta M \frac{\partial \Omega_H}{\partial M}) \nonumber \\
   &= 2 E_{SF}-\frac{S^2}{4 M^3} \, .
\end{align}
Thus, in order that $\delta^2 A \geq 0$, the spin self-force energy must satisfy
\begin{equation}
    E_{SF} \geq \frac{S^2}{8M} \, .
\label{esf2}
\end{equation}
Again, the lower bound yields the actual spin self-force energy if the process of dropping a spinning ring into a Kerr black hole is a reversible process to second order.

\section{Total Self-Force Energy for a Charged and Spinning Body in Kerr-Newman}
\label{tsfe}

Finally, we consider the general case of a charged, spinning body in the Kerr-Newman metric. If the matter responsible for the charge is the same as the matter responsible for the spin, then the spinning charge would create a magnetic dipole moment, and the asymmetric charge distribution would create an electric quadrupole moment. These electromagnetic multipole moments would greatly complicate our analysis. However, there is no need to require that the charge and spin are produced by the same constituents of the body. In particular, we may assume that the charged matter in the body is distributed in a spherical shell and is nonrotating, whereas another uncharged constituent of the body is a rotating ring as considered in the previous section. Thus, it is consistent for us to assume that there are no electromagnetic multipole moments apart from the charge, $q$, of the body.

The analysis of this case can be done in complete parallel to that of the special cases of \cref{csfe} and \cref{ssfe}. The energy, $\epsilon$ now has contributions from both the charge and the spin, as given by \cref{endef}. At first order, we find 
\begin{equation}
    \epsilon > q \frac{Q r_0}{r_0^2+a^2}+ a S \frac{2 M r_0 -Q^2}{(r_0^2 + a^2)^2} \, ,
\label{en8}
\end{equation}
with equality arbitrarily close to being achieved as $r_0 \to r_+$. It follows that by dropping the body from arbitrarily close to the horizon, we can make $\delta A$ arbitrarily close to $0$.

The finite size correction to $\delta^2 M$ due to the charge is again given by \cref{fsq}. Again, there is no finite size correction due to spin. Thus, we have 
\begin{equation}
\frac{1}{2} \delta^2 M = E_F + E_{SF} = \frac{1}{2} \kappa q^2 + E_{SF} \, .
\end{equation}
Thus, the second order change in area is given by
\begin{align}
    \frac{\kappa}{8\pi}\delta^2 A &= \delta^2 M -\delta (\Omega_H)\delta J - \delta(\Phi_H)\delta Q \nonumber \\
    &= \delta^2 M- q^2\frac{r_+}{r_+^2+a^2}-\frac{(S r_+-q a Q)^2}{M(r_+^2+a^2)^2} \nonumber \\
    &= 2E_{SF} + q^2 \frac{r_+ -M}{r_+^2+a^2} -  q^2\frac{r_+}{r_+^2+a^2}-\frac{(S r_+-q a Q)^2}{M(r_+^2+a^2)^2} \, ,
\end{align}
where we used the formula (\ref{kappa}) for $\kappa$.

Thus, the condition $\delta^2 A \geq 0$ gives the following lower bound on the total self-force energy in the general case of a charged and spinning body
\begin{equation}  
    E_{SF} \geq \frac{q^2 M^2(r_+^2+a^2)+(S r_+-q a Q)^2}{2M(r_+^2+a^2)^2} \,.
\label{totsf}
\end{equation}
Again, the lower bound would be the exact expression for self-force energy if the dropping process from the horizon is reversible to second order.

As can be seen from \cref{totsf}, the self-force energy is always positive, corresponding to an overall repulsive self-force. It is interesting that ``cross-terms'' in $S$ and $q$ arise in $E_{SF}$ when both $a \neq 0$ and $Q \neq 0$. This is not unreasonable, because when $Q \neq 0$, the charge perturbs the metric to first order, which can then interact with the spin.

\bigskip

\noindent
{\bf Acknowledgements} We wish to thank Jon Sorce and Abe Harte for reading the manuscript and providing comments. This research was supported in part by NSF grant PHY18-04216 to the University of Chicago.

\appendix
\section{Riemann Tensor on the Symmetry Axis}
\label{app}

In this Appendix, we establish properties of the Riemann tensor on the symmetry axis of a general stationary, axisymmetric spacetime. In the case where the spacetime is a solution of the source free Einstein-Maxwell equations, we will show that the ``quadrupole torque term'' of \cref{eoms} vanishes.

Let $p$ be a point on the symmetry axis. It is convenient to introduce a null tetrad $n^a,l^a, m^a,\bar{m}^a$  at $p$, where $l^a$ and $n^a$ are real, future directed null vectors tangent to the symmetry axis and $m^a$ is a complex spacelike vector orthogonal to the symmetry axis. As usual, the null tetrad is normalized so that
\begin{equation}
    -l^a n_a=m_a\Bar{m}^a=1 \, ,
\end{equation}
and all other inner products vanish.

We can expand $R_{abcd}$ in the basis defined by this null tetrad.
Under the rotations $\phi \to \phi + \chi$ associated with the axisymmetry of the spacetime, $l^a$ and $n^a$ remain invariant, but $m^a$ changes by a phase, $m^a \rightarrow e^{i \chi} m^a$. Since the Riemann tensor must be invariant under these rotations, this implies that an equal number of $m^a$'s and $\Bar{m}^a$'s must occur in each term in this basis expansion. Taking account of this fact together with the Riemann symmetries $R_{abcd}=-R_{bacd}$ and $R_{abcd}=R_{cdab}$ as well as the reality of the Riemann tensor, we see that
the possible terms that can occur in its basis expansion are
\begin{align}
    R_{abcd}&=A l_{[a}n_{b]}l_{[c}n_{d]} + B m_{[a}\Bar{m}_{b]}m_{[c}\Bar{m}_{d]} + i C\left(l_{[a}n_{b]}m_{[c}\Bar{m}_{d]} +m_{[a}\Bar{m}_{b]}l_{[c}n_{d]} \right) \nonumber \\ 
    &+ D \left( l_{[a}m_{b]}n_{[c}\Bar{m}_{d]} + n_{[a}\Bar{m}_{b]}l_{[c}m_{d]} \right) + \bar{D} \left( l_{[a}\Bar{m}_{b]}n_{[c}m_{d]} + n_{[a}m_{b]}l_{[c}\Bar{m}_{d]} \right) \nonumber \\
    & + E \left(l_{[a}m_{b]}l_{[c}\Bar{m}_{d]} + l_{[a}\Bar{m}_{b]}l_{[c}m_{d]} \right) + F \left(n_{[a}m_{b]}n_{[c}\Bar{m}_{d]} + n_{[a}\Bar{m}_{b]}n_{[c}m_{d]} \right) \, ,
\label{riem1}
\end{align}
where the basis expansion coefficients $A,B,C,E,F$ are real and $D$ is complex. The additional Riemann symmetry $R_{[abc]d}=0$ yields the further condition,
$D - \bar{D} = iC$,
thus reducing the general form of the Riemann tensor on the symmetry axis to 6 real parameters. It can be seen that there is a 4-parameter freedom in the components of the Ricci tensor and a 2-parameter freedom in the components of the Weyl tensor. The Weyl tensor has ``type D'' form on the symmetry axis, with repeated principal null directions $l^a$ and $n^a$.

We now make the additional assumption that the spacetime metric satisfies Einstein's equation with electromagnetic stress-energy source
\begin{equation}
    R_{ab}-\frac{1}{2}g_{ab}R=8 \pi T^{\rm EM}_{ab} \, ,
    \label{efe}
\end{equation}
where
\begin{equation}
T^{\rm EM}_{ab}=\frac{1}{4\pi}\left[F_{ac}{F^c}_b-\frac{1}{4}g_{ab}F_{cd}F^{cd} \right] \, .
    \label{mfe}
\end{equation}
We can expand $F_{ab}$ in our null tetrad basis. Axisymmetry of $F_{ab}$ again implies that an equal number of $m^a$'s and $\Bar{m}^a$'s must occur in each term in this basis expansion. Since $F_{ab} = - F_{ba}$, this restricts $F_{ab}$ on the symmetry axis to the form
\begin{equation}
    F_{ab}=G l_{[a}n_{b]}+iH m_{[a}\Bar{m}_{b]} \, . 
\label{mxwlexp}
\end{equation}
where $G$ and $H$ are real.
It follows immediately that
\begin{equation}
l_{[a} F_{b]c}l^c = 0 \, , \quad \quad  n_{[a} F_{b]c}n^c = 0 \, ,
\label{pnd}
\end{equation}
i.e., $l^a$ and $n^a$ are principal null directions of $F_{ab}$. It then follows immediately from \cref{mfe} and \cref{pnd} that
\begin{equation}
T^{\rm EM}_{ab} l^a l^b = T^{\rm EM}_{ab} n^a n^b = 0 \,.
\end{equation}
The Einstein field equation (\ref{efe}) then implies
\begin{equation}
R_{ab} l^a l^b = R_{ab} n^a n^b = 0 \, .
\end{equation}
This, in turn, implies that the coefficients $E$ and $F$ in \cref{riem1} must vanish
\begin{equation}
E=F=0 \, .
\end{equation}
Finally, the trace of \cref{efe} yields $R=0$, which implies that $D + \bar{D}=-(A+B)/2$. 

The quadrupole tensor $J_{abcd}$ of an axisymmetric body has the symmetries of the Riemann tensor, and thus has a basis expansion of the same general form as \cref{riem1}, i.e.,
\begin{align}
    J_{abcd}&=I l_{[a}n_{b]}l_{[c}n_{d]} + J m_{[a}\Bar{m}_{b]}m_{[c}\Bar{m}_{d]} + i K\left(l_{[a}n_{b]}m_{[c}\Bar{m}_{d]} +m_{[a}\Bar{m}_{b]}l_{[c}n_{d]} \right) \nonumber \\ 
    &+ L \left( l_{[a}m_{b]}n_{[c}\Bar{m}_{d]} + n_{[a}\Bar{m}_{b]}l_{[c}m_{d]} \right) + \bar{L} \left( l_{[a}\Bar{m}_{b]}n_{[c}m_{d]} + n_{[a}m_{b]}l_{[c}\Bar{m}_{d]} \right) \nonumber \\
    & + M \left(l_{[a}m_{b]}l_{[c}\Bar{m}_{d]} + l_{[a}\Bar{m}_{b]}l_{[c}m_{d]} \right) + N \left(n_{[a}m_{b]}n_{[c}\Bar{m}_{d]} + n_{[a}\Bar{m}_{b]}n_{[c}m_{d]} \right) \, ,
\label{j1}
\end{align}
with $L- \bar{L} = i K$. One can now calculate the quantity $J^{abcd} {R_{abc}}^e$ directly from the formulas (\ref{riem1}) and (\ref{j1}). Using the fact that $E=F=0$ in the Riemann expansion, one may verify that the result is symmetric in $d$ and $e$. Thus, $J^{abc[d} {R_{abc}}^{e]} = 0$, i.e., the torque term in \cref{eoms} vanishes.

\bibliographystyle{JHEP}

\end{document}